\documentclass[useAMS,usenatbib,usedcolumn]{mn2e}

\usepackage{graphicx} 
\usepackage{times}

\title[Flux density measurements of GPS candidate pulsars at 610 MHz]{Flux density measurements of GPS candidate pulsars at 610 MHz using interferometric imaging technique}
\author[M. Dembska et al.]{M. Dembska$^{1}$\thanks{e-mail: marta.dembska@dlr.de}, R. Basu$^{2,3}$, J. Kijak$^{2}$ and W. Lewandowski$^{2}$ \\
$^{1}$ German Aerospace Center, Institute for Space Systems, Robert Hooke Str. 7, D-28359 Bremen, Germany\\
$^{2}$ Institute of Astronomy, University of Zielona G\'ora, Lubuska 2, 65-265 Zielona G\'ora, Poland \\
$^{3}$ National Centre for Radio Astrophysics, Pune University Campus, Postbag 3, India 411007\\
}
\begin{document}

\date{Accepted\ldots Received\ldots ; in original form\ldots}


\maketitle

\label{firstpage}

\begin{abstract}
We conducted radio interferometric observations of six pulsars at 610 MHz using the Giant Metrewave Radio Telescope (GMRT). All these objects were claimed or suspected to be the gigahertz-peaked spectra (GPS) pulsars. For a half of the sources in our sample the interferometric imaging provides the only means to estimate their flux at 610 MHz due to a strong pulse scatter-broadening. In our case, these pulsars have very high dispersion measure values and we present their spectra containing for the first time low-frequency measurements. The remaining three pulsars were observed at low frequencies using the conventional pulsar flux measurement method. The interferometric imaging technique allowed us to re-examine their fluxes at 610 MHz. We were able to confirm the GPS feature in the PSR B1823$-$13 spectrum and select a GPS candidate pulsar. These results clearly demonstrate that the interferometric imaging technique can be successfully applied to estimate flux density of pulsars even in the presence of strong scattering. \end{abstract}


\section{Introduction}
In the case of most pulsars, their observed radio spectra can be described using a power law with a negative spectral index of $-$1.8 or (for a small fraction of sources) two power laws with spectral indices of $-$0.9 and $-$2.2 with a break frequency $\nu_b$ on average of 1.5 GHz \citep{maron2000}. Some pulsars also exhibit a low-frequency turnover in their spectra~\citep{sieber, lorimer1995}. A spectrum of that kind is characterized by a positive spectral index below a peak frequency $\nu_p$ of about 100 MHz (with a few exceptions when the spectrum peaks at frequencies up to several hundred MHz). However, \citet{kijak2011b} pointed out a small sample of pulsars that peak around 1 GHz and above. Such an object, called the gigahertz-peaked spectrum (GPS) pulsar, is described as a relatively young source that has a high dispersion measure (DM) and usually adjoins a dense, sometimes extreme vicinity. This suggests that the GPS in pulsars might be caused by either the conditions around neutron stars or the physical properties of the interstellar medium. 

The strongest argument for environmental origin of the high-frequency turnover in radio pulsars spectra is the evolution of PSR~B1259$-$63 spectrum. \citet{kijak2011a} showed that the spectrum of the pulsar at the various orbital phases exhibits both a shape and a peak frequency evolution due to the orbital motion of the pulsar around its companion Be star LS 2883 on a very elliptical orbit. The PSR~B1259$-$63 spectrum demonstrates a strong similarity with the gigahertz-peaked spectra, especially when the pulsar in its motion gets closer to its companion star. \citet{kijak2011a} proposed two effects which can be responsible for the observed variations, the free-free absorption in the stellar wind and the cyclotron resonance in the magnetic field associated with the disk of Be star. Both these processes assume the absorption to be caused by external factors, like in the cases of the isolated GPS pulsars~\citep{kijak2011a}. 

\citet{kijak2013} studied the radio spectra of two magnetars PSRs J1550$-$5418 and J1622$-$4950 and in both cases their radio spectra clearly peak at the frequencies of a few GHz. Both these magnetars are associated with supernova remnants and hence surrounded by ionized gas which can be responsible for the free-free absorption of the radio waves. The authors concluded that the GPS feature in radio magnetars spectra can be of environmental origin, in the same way as it occurs in the vicinity of GPS pulsars. 

Pulsars with a high-frequency turnover in their spectra have represented the smallest group of the radio pulsar spectra types. However, \citet{bates2013} estimated that the number of such sources may constitute up to 10\% of the whole pulsar population. The sample of GPS pulsars was extended to include PSR J2007+2722, whose flux density measurements were presented by~\citet{einstein}. Recently,~\citet{dembska2014} reported two newly-identified GPS pulsars. One of them,  PSR B1740+1000, is the first low-DM pulsar that exhibits the gigahertz-peaked spectrum. This case, along with the GPS phenomenon in radio magnetars, led the authors to conclude that the GPS candidate selection criteria need to be revisited. In future searches for new GPS pulsars, the presence of interesting (or extreme) environments, instead of the high DM, could play a crucial role in the source selection process.

\citet{dembska2014} also pointed out that the small number of the currently known GPS pulsars may be the result of our limited knowledge of pulsar spectra in general, especially below 1 GHz. The authors outlined the need for a more extensive sample of GPS sources to establish a plausible statistics about those objects. However, in the cases of some GPS candidate pulsars the standard pulsar flux measurement methods are affected by strong scattering at low frequencies. The phenomenon causes the pulse profiles to become broader, i.e. pulses attain roughly exponentially decaying scattering tail. It has been shown that the characteristic broadening of the pulse, $\tau_{sc}$, depends on both the observing frequency, as well as DM (the empirical relation was given by~\citeauthor{bhat2004}~\citeyear{bhat2004}). Recent results on scattering were discussed by~\citet{lewandowski2013} in their analysis of 45 pulsars, based on the Giant Metrewave Radio Telescope (GMRT) and the Effelsberg Radio Telescope observations. Since the scattering becomes stronger at lower frequencies, for a given pulsar the flux becomes increasingly underestimated then. For high-DM pulsars at low frequencies, when the scattering time is greater than the pulsar period by a significant factor, one will see no pulsed emission. Thus, the flux density measurements required to construct radio pulsar spectra using ``traditional" methods can be difficult or sometimes impossible to conduct. For these cases the only way to determine the pulsar flux is using the interferometric imaging techniques  (see for example \citeauthor{kouwenhoven}~\citeyear{kouwenhoven}).

The interferometric measurements of pulsar fluxes at both 325 MHz and 610 MHz using the GMRT have been demonstrated in~\citet{basu1} and \citet{basu2}. The imaging techniques provide a superior alternative to the standard flux measurements, especially in our studies since the sources we selected for observations are high-DM pulsars. For some of them the imaging techniques are the only secure means to estimate their flux. There is at least two reasons for employing imaging techniques. Firstly, flux calibration in an interferometer is more robust due to the baseline lying at zero level thereby reducing errors made during the baseline subtraction of a normal pulsar observation. Secondly, the instrumental and atmospheric gain fluctuations on very short time scales can be corrected using self-calibration of the interferometric data. The corrections are determined by flux densities of constant and bright background sources in the field and hence would not be affected by the pulse variation of the relatively weak pulsar at a field center. 

In this paper we present flux measurements of six pulsars observed at the 610 MHz frequency band of the GMRT using interferometric imaging technique. The sources selected for our studies are GPS pulsars or GPS candidates. We chose the 610 MHz band for these studies, firstly, because the frequency is low enough to estimate whether a given object is indeed a GPS pulsar, secondly, due to the higher probability of detecting GPS pulsars at 610 MHz than at lower frequencies as a result of their inverted spectra at sub-GHz frequencies and finally, to avoid the RFI and other systematic effects that are prominent at lower frequencies. This analysis allowed us to inspect flux densities of some objects and select a strong GPS candidate pulsar. We were able to confirm that the the interferometric imaging technique can be successfully applied to estimate flux density of pulsars. 

\section{Observations and data analysis}

\setcounter{table}{0}
\begin{table}
\resizebox{\hsize}{!}{
\begin{minipage}{85mm}
\caption{The list of sources and their parameters, where $DM$ -- dispersion measure, $P$ is period and $\tau_{sc}$ denotes scattering time.}
\centering
\begin{tabular}{@{}l c D{.}{.}{3.1} l D{.}{.}{1.2} c@{}}
\hline
& & & & & \\
Pulsar & $DM$ & \multicolumn{1}{l}{Age} & \multicolumn{1}{c}{$P$} & \multicolumn{1}{c}{$\tau_{sc}$} & Associations \\
    & $\rm{\left(\frac{pc}{cm^{3}}\right)}$ & \multicolumn{1}{l}{(kyr)}  & \multicolumn{1}{c}{(sec)} & \multicolumn{1}{c}{(sec)} &\\
\hline
B1750$-$24   & 672  & 593 &  0.528 & 2.38  & \\

B1815$-$14  & 622 & 2270 & 0.291 & 0.457  & \\

B1849+00  &   787   &  356 &  2.18 & 6.19  & \\

& & & & & \\

B1800$-$21  &     234     &  15.8 &   0.133 &  0.039  & 1, 2, 3\\

B1822$-$14 & 357 & 195 & 0.143 &  0.279 & 1\\

B1823$-$13  &    231   &  21.4 & 0.101 & 0.0337 & 1, 2 \\
\hline
& & & & & \\
\multicolumn{6}{l}{$1$ -- HESS (High Energy Stereoscopic System)}\\
\multicolumn{6}{l}{$2$ -- X-PWN (X-ray pulsar wind nebula), $3$ -- SNR (supernova remnant)}
\end{tabular}
\label{tabGMRT1}
\end{minipage}
}
\end{table}

\begin{figure}
   \begin{flushleft}
   \includegraphics[width=8.5cm]{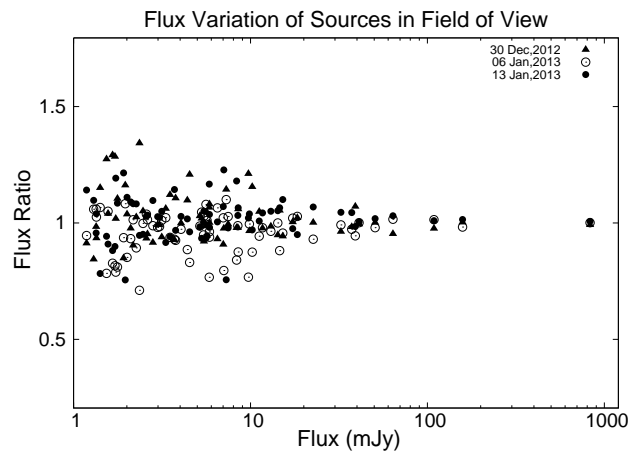}
   \caption{The figure shows the variation of the flux value of the surrounding sources in the field of  view across the three observing runs. The strong sources in the fields with the pulsars PSRs B1750$-$24 and  B1849+00 at the field centers were identified and their average flux across the three sessions were calculated. The ratio of the flux of each source with respect to the average flux and as a function of the average flux is plotted in the figure, with each session marked with different symbols. The ratios are scattered around unity with the noise in the scatter decreasing with the increasing flux levels. This demonstrates the analysis process to be correct and consistent for all the three observing sessions.}
  \label{flux_ratio}
   \end{flushleft}
\end{figure}
We recorded the interferometric data on six pulsars using the GMRT that is located near Pune, India and consists of an array of 30 distinct dishes, each with a diameter of 45 meters, and a total of 435 baselines. The dishes are spread out over a region of $\sim$ 27 km and roughly resemble a Y-shaped array. The data were recorded at the 610 MHz frequency band with a bandwidth of 33 MHz spread over 256 frequency channels. All the six sources were observed on three separate dates: 30 December 2012, 6 January 2013 and 13 January 2013, each observation separated by a week, to account for variations in the pulsar flux over long time scales.

\begin{table}
\resizebox{\hsize}{!}{
\begin{minipage}{85mm}
\caption{Flux density measurements resulted from the interferometric imaging observations in three observing sessions ($S_1$, $S_2$ and $S_3$ respectively) along with the uncertainties which include calibration errors, the rms noise in the maps and fitting errors. $\left<S\right>$ denotes the weighted mean of all results (presented along with its uncertainty) for a given pulsar.}
\centering
\begin{tabular}{@{}l  D{,}{\pm}{3.3}  D{,}{\pm}{3.3}  D{,}{\pm}{4.4}  D{,}{\pm}{4.4}@{}}
\hline
& & & &  \\
Pulsar & \multicolumn{1}{c}{$S_1$} & \multicolumn{1}{c}{$S_2$} & \multicolumn{1}{c}{$S_3$} & \multicolumn{1}{c}{$\left<S\right>$}\\
& \multicolumn{1}{c}{(mJy)} & \multicolumn{1}{c}{(mJy)} & \multicolumn{1}{c}{(mJy)} &\multicolumn{1}{c}{(mJy)}\\
\hline
B1750$-$24 &3.65,0.28& 3.90,0.25 & 3.98,0.25 & 3.9,0.3\\

B1815$-$14 & 24.5,1.8 & 24.6,1.5 & 26.1,1.5 & 25.1,1.6\\

B1849+00 & 15.2,1.1 & 15.5,1.0 & 16.1,1.0 & 15.39,0.95\\

& & & &  \\

B1800$-$21 &8.22,0.60 & 6.75,0.42 & 7.82,0.48 & 7.4,0.9  \\

B1822$-$14 & 3.31,0.36 & 3.43,0.35 & 4.18,0.38 & 3.5,0.6\\

B1823$-$13 & 3.51,0.26 & 3.42,0.22 & 3.53,0.23 & 3.5,0.2\\

\hline
& & & &  \\
\end{tabular}
\label{tabGMRT2}
\end{minipage}
}
\end{table}

The observations were carried out using standard schemes where strategically placed calibrators were interspersed with the sources. We recorded flux calibrators 3C48 and 3C286 before and after each observing run for 10 min. The antenna gain variations were determined using a phase calibrator observed every half hour. The sources in our sample are nearby in terms of their position in the sky  and the phase calibrator 1822$-$096 was used for all the sources to carry out amplitude and phase calibrations as well as bandpass  calibration to correct for variation across frequency band. Each of the pulsars was observed for 10 minutes at a time and six such observations on each source were well spaced out during each of the observing sessions which lasted for about eight hours. We recorded one hour of data on each pulsar during each observing run with all the sources having a wide coverage in the u-v plane due to our strategy of spacing out their pointings. 

The data reduction was done using the Astronomical Image Processing System (AIPS) where we used standard  techniques of RFI flagging, calibration and imaging. The flux scales were determined using the measurements  of \citet{baars} with latest corrections by Perley et al. 2010 (in AIPS). All the pulsar fields during each observing run was passed through multiple rounds of self-calibration and imaging to improve noise characteristics in the maps. The following checks were performed to test the fidelity of the images: \\
1. The flux of the source 1822$-$096 (phase calibrator) was determined for each of the three observing sessions using both the flux calibrators 3C48 and 3C286. The flux value for each of the sessions were 6.3$\pm$0.5 on 30 December 2012, 6.2$\pm$0.5 on 6 January 2013 and 6.3$\pm$0.5 on 13 January 2013. This shows that the various independent calibrations applied during each separate observations and also different flux calibrators gave identical results demonstrating the technique to be correct. In addition the correctness of the flux scale was demonstrated by comparing the flux value of the phase calibrator to the expected value (between 325 MHz and 1.4 GHz) which yielded consistent results.\\
2. We performed an additional test for determining the correctness of the analysis techniques. The same fields were observed on multiple days with independent calibrations for the system. Each of the pulsar fields imaged, with the pulsar at a field center, had a half power radius of $\sim$ 40 arcmin with a large number of sources \mbox{($>$ 100)} in each field of view. If an analysis is correct then the flux of all these sources, in an ideal scenario, should be identical for all the observations. However, under realistic conditions one expects some noise across the different observing runs. To demonstrate this we used two (out of six) pulsar fields and determined the flux values of all the strong sources (flux greater than 10 times the rms noise in a map) for all the three observing runs. An average flux for each source across the three observing runs was calculated and the ratio of the flux to average flux was plotted as a function of the source flux value (see Figure~\ref{flux_ratio}). As is clear from the figure the flux ratios are concentrated around unity with the noise increasing with decreasing flux level of a source. This is exactly as per our expectations and clearly demonstrates the correctness of our calibration and analysis techniques across the three observing sessions.

\section{Results}

Table~\ref{tabGMRT1} gives the list of sources with some of their basic parameters (DM, age and period). All these objects are either confirmed or candidate GPS sources. As it is clear from the table, the sample was subdivided into two categories. We have made rough estimates of the scattering timescales using either observational data~\citep{lewandowski2013} or the predictions derived from a single thin screen model (using the scatter time frequency  scaling index $\alpha=4$). For the first group of pulsars, PSRs B1750$-$24, B1815$-$14 and B1849+00, the pulse broadening at 610 MHz is large, hence the interferometric imaging provides the only mean to estimate flux for these sources. The flux density of the remaining pulsars, PSRs B1800$-$21, B1822$-$14 and B1823$-$13, was measured at low frequencies by conventional pulsar observations \citep{lorimer1995, kijak2007, kijak2011b}, however since the predicted (or measured) scatter time estimates are a significant fraction of the pulsar periods, there is a possibility for the flux densities measured that way to be underestimated. Additionally, these three sources have counterparts in the High Energy Stereoscopic System (HESS) observations with indications of pulsar wind nebula (PWN) around them. \citet{kijak2011a} and \citet{kijak2011b} reported PSRs B1822$-$14 and B1823$-$13 as GPS pulsars, and PSR B1800$-$21 is considered as a GPS candidate. We included these pulsars in our studies to re-examine their flux values at 610 MHz where some profile broadening due to scattering is also present.

\begin{figure}
   \begin{center}
   \includegraphics[width=8.2cm]{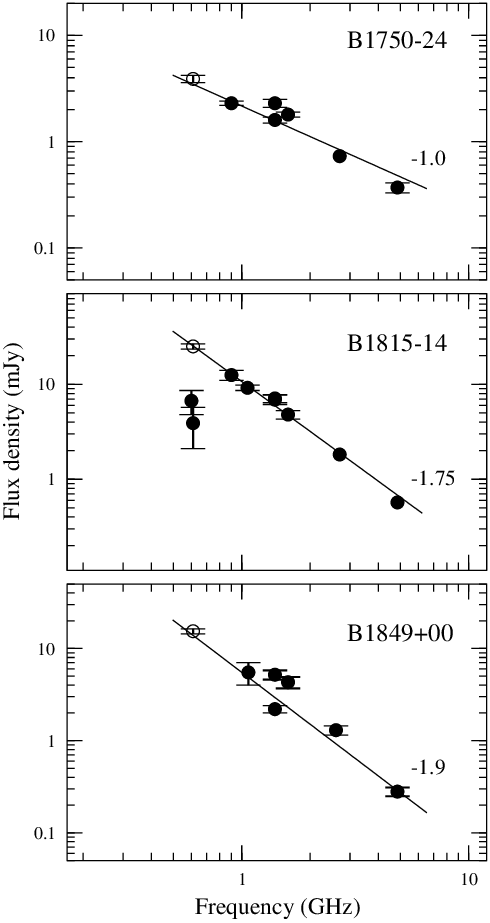}
   \caption{The spectra of PSRs B1750$-$24, B1815$-$14 and B1849+00, pulsars with high DMs (see Tab. \ref{tabGMRT1}). Open circles denote the GMRT interferometric observations. Measurements marked with black dots are taken from literature \citep[and references therein]{maron2000, kijak2007, kijak2011b} and the ATNF pulsar catalogue. The straight lines represent our fits to the data using power-law function (in the case of PSR B1815$-$14 the flux measurements at 610 MHz marked with black dots were excluded from the fitting procedure). The power indices resulted from the fitting procedure are given on each panel, for more details see Tab.~\ref{fits}.}
  \label{highDM}
    \end{center}
\end{figure}

For the six pulsars we constructed radio spectra, combining flux density measurements from the literature~\citep[and references therein]{lorimer1995, maron2000, kijak2007, kijak2011b} and the ATNF (Australian Telescope National Facility) pulsar catalogue\footnote{http://www.atnf.csiro.au/research/pulsar/psrcat/ \citep{atnf}}  together with the new results shown in Table~\ref{tabGMRT2} (see online material for the maps of the pulsars that were used to estimate their flux density at 610 MHz). The spectra are presented in groups, depending on their morphological properties. Our studies show that for the high-DM pulsars, namely PSRs B1750$-$24, B1815$-$14 and B1849+00, their spectra resemble a simple power law. We were able to confirm the GPS feature in the spectrum of PSR B1823$-$13 and pointed out two sources, PSRs B1822$-$14 and B1800$-$21, suspected to be gigahertz-peaked spectra pulsars which require further investigation. The fits presented in the paper were obtained by the implementation of the nonlinear least-squares Levenberg-Marquardt algorithm.The results of the fitting procedure are given in Tab. \ref{fits}.\\

\textbf{PSRs B1750$-$24, B1815$-$14 and B1849+00} have very high DMs in excess of 600~pc~cm$^{-3}$. Their spectra are shown in Figure~\ref{highDM}. These objects are characterized by a significant pulse broadening due to interstellar scattering (of the order of several periods). As mentioned above, the sources were suspected to be GPS pulsars but our analysis suggest that their spectra can be fit by a power law. 

In the case of PSR B1815$-$14 the interferometric imaging technique was used to inspect its flux density measurements at 610 MHz due to a strong pulse scatter-broadening~\citep[over a full pulse period, see][]{lewandowski2013}. Previous observations suggested that the pulsar spectrum shows a turnover feature. Our new result, accompanied by a large scatter time estimate, suggests a significant underestimation in the previous flux measurements for this pulsar at 610~MHz. 

\setcounter{table}{2}
\begin{table}
\resizebox{\hsize}{!}{
\begin{minipage}{80mm}
\caption{Fitted parameters to data of four pulsars, PSRs B1750$-$24, B1815$-$14 and B1849+00 (using a power-law, where $\xi$ is a power index) and B1823$-$13 (using the same function (1) as in~\citet{kijak2011a}). The implementation of the nonlinear least-squares Levenberg-Marquardt algorithm was used to perform the fitting procedure. The parameters are given with a reduced $\chi^2$. For more details on data excluded from the fitting procedure see Figs \ref{highDM} and \ref{GPS}.}
\centering
\begin{tabular}{@{}l l   D{,}{.}{1.3}@{} }
\hline
& &  \\
Pulsar & Fitted parameters & \multicolumn{1}{r}{$\chi^2$} \\
\hline
B1750$-$24 & $\xi=-1.0 \pm 0.14$ & 7,5  \\
B1815$-$14 & $\xi=-1.75 \pm 0.07$ & 0,95  \\
B1822$-$14 & $\xi=-0.48\pm 0.08$ & 2,5 \\
B1849+00 & $\xi=-1.9 \pm 0.2$ & 9,6 \\
& &  \\
B1823$-$13 & $a = -0.95\pm 0.18$, $b = 0.28 \pm 0.11$&  3,4 \\
&$c = 0.61\pm 0.03$  & \\
\hline
 & &  \\
\end{tabular}
\label{fits}
\end{minipage}
}
\end{table}

The classification of the PSRs B1750$-$24 and B1849+00 spectra was not possible before due to very limited flux density measurements at frequencies below 1 GHz. It seems clear that the spectrum of PSR~B1849+00 one is a typical steep pulsar spectrum. The PSR~B1750$-$24 spectrum can also be described by a single power law but at the same time it is relatively flatter than a usual pulsar spectrum (a spectral index of $-1.0$; the fits and the resulting spectral indices are presented in Fig.~\ref{highDM}).\\

\textbf{PSR~B1823$\mathbf{-}$13}, whose spectrum is presented in Figure~\ref{GPS}, was classified as a GPS pulsar by~\citet{kijak2011b}. The object was included in our sample to re-examine its flux density measurement at 610 MHz. It is noteworthy that new measurements indicate a larger flux density than previous GMRT observations which were carried out using the instrument in phased array mode, thus standard pulsar flux measurement methods were applied~\citep{kijak2011b}. Even if one disregards the earlier measurements and uses only the interferometrically derived flux, the spectrum continues to exhibit a GPS feature. However, it seems clear that this object requires further observations at frequencies below 600 MHz which we plan to perform in future projects. \\

\begin{figure}
   \begin{flushleft}
   \includegraphics[width=8.2cm]{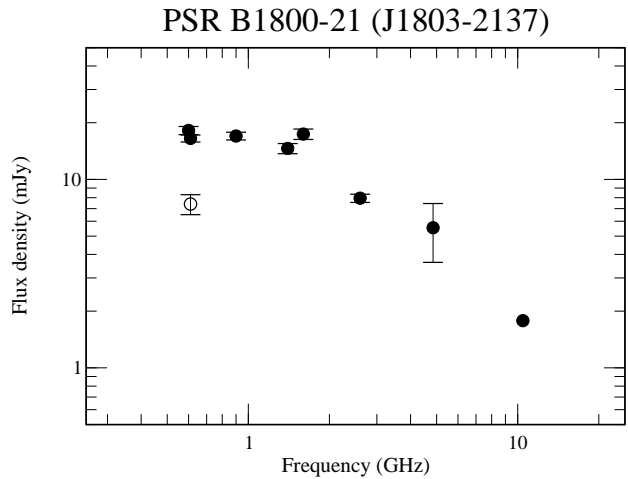}
   \caption{The spectrum of the candidate for GPS pulsar, PSR B1800$-$21. Open circle denotes the GMRT interferometric observations. Measurements marked with black dots are taken from literature \citep[and references therein]{maron2000, kijak2011b} and the ATNF pulsar catalogue.}
  \label{1800}
   \end{flushleft}
\end{figure}

\textbf{PSR~B1800$-$21}, whose spectrum is presented in Fig.~\ref{1800}, can be treated as a new, very promising GPS candidate pulsar. It is a young, Vela-like pulsar, associated with a supernova remnant~\citep{kijak2011b}. Previously the spectrum seemed to be flat at low frequencies but our estimate, which indicates a much smaller flux density than in the previous measurements at 610 MHz, suggests a positive spectral index in the low-frequency range. New results clearly imply that the earlier flux measurements should be verified.

\begin{figure}
   \begin{flushleft}
   \includegraphics[width=8cm]{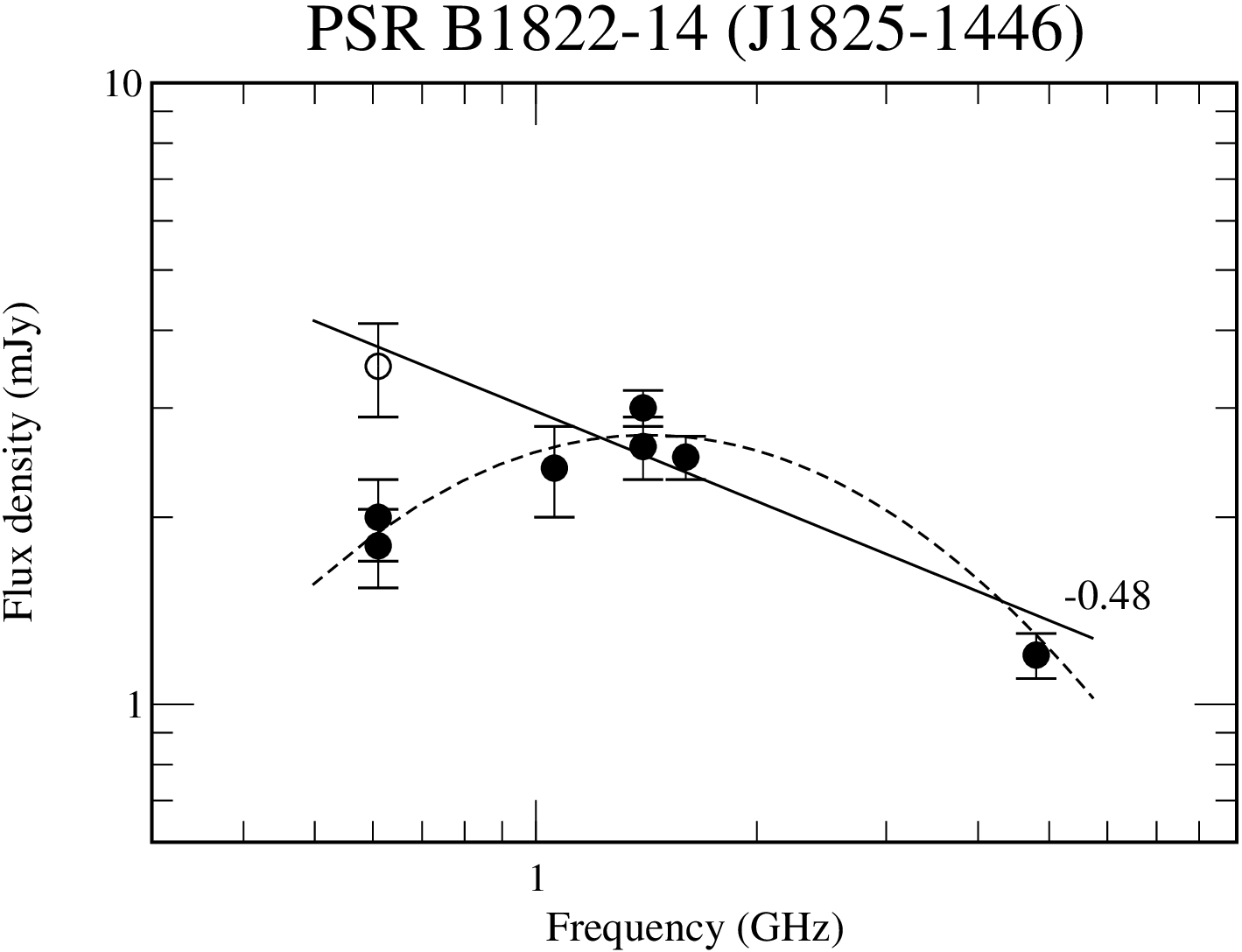}
   \caption{The spectrum of  the GPS pulsar PSR B1822$-$14. Open circle denotes the GMRT interferometric observations, whereas black dots denote data
from literature \citep[and references therein]{maron2000,kijak2007,kijak2011a} and the ATNF pulsar catalogue. The dashed curve represents the fit to the data made by~\citet[see their Fig. 1, a point marked with an open circle was not included]{kijak2011a}. We present our power-law fit (the straight line) to the data, excluding two measurements at 610 MHz marked with black dots. The fitted power index is given with the fit, for more details on the fitting results see Tab.~\ref{fits}.}
  \label{1822}
   \end{flushleft}
\end{figure}

\textbf{PSR~B1822$-$14} is the last pulsar in our sample. Its spectrum is presented in Figure~\ref{1822}, Similar to the case of PSR~B1823$-$13, the flux density measured using the interferometric imaging techniques is greater than the values from standard method. PSR~B1822$-$14 was identified as a GPS pulsar~\citep{kijak2011b}, however the interferometric measurements suggest otherwise -- the spectrum of the pulsar seems to be a power law (with spectral index of $-0.48$) when including new measurements which may indicate that the earlier measurements were affected by interstellar scattering. This pulsar definitely needs further investigation and possibly additional measurements at lower frequencies.

\section{Discussion}
The interferometric imaging technique allowed us to measure flux density for six pulsars with the GMRT at 610 MHz. The frequency was chosen to verify whether the objects are GPS pulsars. In our studies we included three pulsars which had previous flux measurements at 610 MHz (obtained by using the standard techniques) in order to re-examine these estimates.

In the case of the remaining three sources, due to a strong pulse broadening, the interferometric observations are the only way to estimate their flux at this frequency (see the scatter time estimates in Tab. \ref{tabGMRT1}). We inspected the spectra of these three pulsars with very high-DMs ($>$600 pc cm$^{-3}$). PSRs 1750$-$24, 1815$-$14 and B1849+00 were suspected to be GPS sources but with the addition of new measurements their spectra appear to resemble a typical single power-law pulsar spectrum.  We can also confirm the GPS feature in the PSR~B1823$-$13 spectrum and select a new strong GPS candidate: PSR~1800$-$21. Its spectrum as well as the spectrum of PSR B1822$-$14, require however some further investigation. 

\begin{figure}
   \begin{flushleft}
   \includegraphics[width=8.2cm]{1823-13.eps}
   \caption{The spectrum of the GPS pulsar PSR B1823$-$13. Open circle denotes the GMRT interferometric observations, whereas black dots denote data
from literature \citep[and references therein]{maron2000,kijak2007,kijak2011a} and the ATNF pulsar catalogue. The curve represents our fits to the data using the same function (1) as in~\citet{kijak2011a}. The measurement at 610 MHz marked with the black dot was excluded from the fitting procedure. For fitted parameters see Tab.~\ref{fits}}
  \label{GPS}
   \end{flushleft}
\end{figure}

\subsection{The interstellar scintillations and scattering}
The variations in flux density values resulted when applying different observing techniques can be caused by interstellar scintillations. The pulsars in our sample have DMs ranging from $\sim$200 to $\sim$800 pc cm$^{-3}$ which means that any scintillations will occur in a strong scintillation regime. Hence we have to take into consideration both diffractive and refractive scintillations. As the diffractive time scales, $\Delta t_{\rm{DISS}}\propto f^{1.2} d^{-0.6}$, where $d$ denotes distance, in a case of high-DM and hence distant sources should be up to a few minutes at the observing frequency $f$ of 610~MHz, thereby a single 1~hour observing run should be enough to averaged-out any variations of the measured flux density. The discrepancies with the earlier flux density measurements (if present) can be at least partially caused by the refractive interstellar scintillations, as the refractive time scales, $\Delta t_{\rm{RISS}}\propto f^{-2.2} d^{1.6}$, may vary from days to months. However, the refractive modulation index $\Delta m_{\rm{RISS}}\propto f^{0.57} d^{-0.37}$ for high-DM pulsars should be relatively small at the frequency of 610~MHz. To account for refractive scintillations influence, we performed our flux measurements on three epochs (separated by no less than a week). For the purpose of constructing pulsar spectra we are using an average of values obtained from these measurements. We have to note however, that in this regard the interferometrically measured flux density values are not affected any more than the ones resulting from the standard observing procedure. Pulsar flux measurements regardless of the method used will be affected by the scintillation-driven flux variation in exactly the same way.

As mentioned in the Introduction, for high-DM pulsars, and especially when the observations are conducted at low observing frequencies, the pulsar profiles can be affected by interstellar scattering to such an extent that it is difficult, or even impossible to perform flux measurement using the standard profile-based method. Strong scattering causes the profiles to attain scattering tails, what may significantly affect the observed profile background level so that the proper baseline required for the standard flux measurement can not be found. Especially prone to such errors will be the cases where the scatter time is comparable to the pulse period, as one will still be able to see a prominent pulse, while the scattering tail will hide the proper baseline level. One has to remember that even a ``moderate'' length of the scattering tail may still affect the profile baseline and cause erroneous results, especially when the observations are affected by a high noise level \citep{lewandowski2013}. In some circumstances the influence of scattering may even change the appearance of the spectrum by mimicking a turnover feature, which may lead to an erroneous identification  of a pulsar as a GPS source, like in the case of PSR B1815$-$14.

Since there is no simple way to account for the scattering-induced baseline level change one should refrain from using the standard method in such cases. Practically, however, it can be hard to judge if  we are dealing with a profile where scattering led to an erroneous flux density estimation. We believe that at least some of the past flux measurements performed for the pulsars from our sample might have been affected by the scattering, which is discussed below. Regardless, to avoid the scattering-induced issues one needs a reliable method of pulsar flux density measurement that will not be affected by scattering.

\subsection{Imaging technique and pulsar flux density}
Interferometric imaging techniques at low radio frequencies have been used to determine flux density of a wide class of astronomical sources and can be easily extended to determine pulsar flux values. The estimated flux of the phase calibrator 1822$-$096 at 610 MHz was identical over the three widely separated observing sessions and using two different flux calibrators 3C48 and 3C286. In addition its flux estimates are in the expected range when compared with the flux densities at 325 MHz and 1.4 GHz (from the Very Large Array calibrator manual). The consistency of our calibrations over the different observing sessions was also demonstrated by the flux ratios of the surrounding background sources in the field of view, which hovered around unity (see Figure~\ref{flux_ratio}). Additional vindication of our results is provided by the flux density measurements of the three pulsars B1750$-$24, B1815$-$14 and B1849+00 which could not be determined at 610 MHz using conventional pulsar flux measurement techniques due to their highly scattered profiles. Our flux estimates at 610 MHz for all the three cases are consistent with the pulsar spectra determined from higher frequencies, where interstellar scattering should not affect the measurement  (see Figure~\ref{highDM}).

We now examine the cases of the three other pulsars B1800$-$21, B1822$-$14 and B1823$-$13 where traditional pulsar flux measurements have been carried out in the past. All of these sources adjoin interesting environments which may cause an additional absorption at frequencies below 1 GHz. \citet{dembska2014} suggested that an influence of such environment on pulsar radio emission can manifest itself as a reverse spectrum with positive spectral index, which made these objects a plausible GPS candidates.

Two of the pulsars, PSRs B1822$-$14 and B1823$-$13, were classified by \citet{kijak2011a} and \citet{kijak2011b} as GPS pulsars. Obviously, one can note the discrepancies between standard and interferometric flux measurements for both sources. The interferometrically measured flux densities are larger than the ones which resulted from standard observations. We know that the pulse profile of PSR B1822$-$14 at 610 MHz shows a significant scattering~\citep{lewandowski2013}, thus the flux density obtained using the standard method is very likely to be underestimated. In the case of  PSR B1823$-$13 the profile that was used by \citet{lorimer1995} to estimate the flux density of the pulsar at 610 MHz was not published. However, the estimated scatter time at this frequency is a significant fraction of the pulsar period (as in the case of PSR B1822$-$14). Hence one cannot rule out the possibility that conventional observations were in this case also affected by scattering. 

\subsection{The case of PSR B1800$-$21}
As we pointed out above, usually the interferometric observations give better estimations of previously underestimated flux density values. Yet, one can note that in the case of PSR B1800$-$21 its interferometrically measured flux value is lower that the one resulted from previous measurements by~\citet{lorimer1995} and~\citet{kijak2011b}. The discrepancy cannot be explained in a simple way and it is not possible to definitely point out the reason behind it. However, one still can propose possible effects causing such a variation. Assuming our latest result to be accurate and precise, there appears a chance that the difference is caused by rather external (i.e. environmental) factors. PSR B1800$-$21 is a young Vela-like pulsar, located at the southwestern edge of the G8.70.1 nebula, a shell-type supernova remnant~\citep{kw1,kw2}. It has a plausible association with the $\gamma$-ray source HESS J1804-216~\citep{ah1,ah2,hig}. All these objects, together with a number of discrete ultracompact HII regions, compose the W30 complex of  peculiar morphology~\citep{fo}, which emits over frequencies ranging from radio to $\gamma$-rays. Thus, it is possible that during our interferometric observations the observed  PSR B1800$-$21 flux was affected by an additional absorption, which manifested itself at lower frequencies. Such transient absorption may appear due to relative motion of the components of the W30 complex. 

One should note that the two standard measurements which yielded higher flux values were separated in time (by more than 10 years) and conducted independently using different observing facilities. Nevertheless, one cannot exclude the influence of the classical pulsar flux measurement procedure, such as non-linearity of a receiver or some non-typical flux calibration issues. Either way, PSR B1800$-$21 definitely requires further investigation. We have already conducted another set of interferometric observations of this source at 610 MHz with the GMRT to verify its flux density. The data is still to be analysed, however preliminary results show that the flux density resulted from our observations conducted on 28 December 2013 is $8.79\pm0.67$ mJy which is in a good agreement with values presented in Tab. \ref{tabGMRT2}. Moreover, we have time allocation for 325~MHz observations of this source -- these observations are to be conducted in January 2015. Consequently, we will be able to re-examine our results and hopefully resolve the discrepancy with the standard method measurements.\\

\section{Summary}
The interferometric imaging technique allowed us to estimate pulsar flux density, making it possible to observe weak or high-DM pulsars at frequencies below 1 GHz and re-examine our previous results. In the cases of objects for which determining their flux at low frequencies using standard pulsar observing methods is not possible (i.e. for a large $\tau_s$ when comparing to pulsar period), it is the only way to estimate flux for these sources. Thus, the interferometric imaging technique provides a superior alternative to the pulsar flux measurements. We believe that the technique will help us to confirm more GPS pulsars in future.\\

Moreover, we used this method to rule out three GPS pulsar candidates and confirm one GPS pulsar, PSR B1823$-$13. We also pointed out two GPS candidates, including a very promising one, PSR 1800$-$14, that require further investigation. Hence, our analysis clearly shows that the interferometric imaging technique can be successfully applied to estimate flux density of pulsars. The method is the only secure way to determine the flux density of high-DM pulsars which are highly scattered at low radio frequencies.

\section*{Acknowledgments}
We thank the staff of the GMRT who have made these observations possible. The GMRT is run by the National Centre for Radio Astrophysics of the Tata Institute of Fundamental Research. This research was partially supported by the grant DEC-2013/09/B/ST9/02177 of the Polish National Science Centre. MD was a scholar within Sub-measure 8.2.2 Regional Innovation Strategies, Measure 8.2 Transfer of knowledge, Priority VIII Regional human resources for the economy Human Capital Operational Programme co-financed by European Social Fund and state budget. We thank M. Jamrozy for this support on the preparation of the observing proposal.

\end{document}